# Phase synchronization of instrumental music signals

Sayan Mukherjee[1a], Sanjay Kumar Palit[2], Santo Banerjee[3], MRK Ariffin[4], D.K.Bhattacharya[5]

[1a] Mathematics Department, Sivanath Sastri College, Kolkata, India
[2] Basic Sciences & Humanities Department, Calcutta Institute of Engineering and Management, Kolkata, India
[3] Institute for Mathematical Research, Universiti Putra Malaysia, Malaysia
[4] INSPEM and Department of Mathematics, Universiti Putra Malaysia, Malaysia
[5] Department of Instrumental Music, Rabindra Bharati University, Kolkata-700050, INDIA

**Abstract.** Signal analysis is one of the finest scientific techniques in communication theory. Some quantitative and qualitative measures describe the pattern of a music signal, vary from one to another. Same musical recital, when played by different instrumentalists, generates different types of music patterns. The reason behind various patterns is the psycho-acoustic measures – Dynamics, Timber, Tonality and Rhythm, varies in each time. However, the psycho-acoustic study of the music signals does not reveal any idea about the similarity between the signals. For such cases, study of synchronization of long-term nonlinear dynamics may provide effective results. In this context, phase synchronization (PS) is one of the measures to show synchronization between two non-identical signals. In fact, it is very critical to investigate any other kind of synchronization for experimental condition, because those are completely non identical signals. Also, there exists equivalence between the phases and the distances of the diagonal line in Recurrence plot (RP) of the signals, which is quantifiable by the recurrence quantification measure $\tau$-recurrence rate. This paper considers two nonlinear music signals based on same raga played by two eminent sitar instrumentalists as two non-identical sources. The psycho-acoustic study shows how the Dynamics, Timber, Tonality and Rhythm vary for the two music signals. Then, long term analysis in the form of phase space reconstruction is performed, which reveals the chaotic phase spaces for both the signals. From the RP of both the phase spaces, $\tau$-recurrence rate is calculated. Finally by the correlation of normalized tau-recurrence rate of their 3D phase spaces and the PS of the two music signals has been established. The numerical results well support the analysis.

## 1. Introduction

In the modern digital world, the theory of communication is an indispensible scientific approach to encrypt the meaning of music. The language of communication defines how computer and our brain recognize the music. In fact, the learning process of music in our brain is explained by different physical measures by the computation of digitized music signal in computer. In this concern, scientists and music experts are trying to extract intelligible information from music in different ways. Whatever may be the way of extraction, the main purpose is to make coherence between music and social environment by some simple informative knowledge which can be easily interpreted.

Also in our busy world, most of the peoples are facing stress and anxiety which have a large negative impact on our society and most of our diseases are originated from psycho-somatic disorder. For this reasons, people are affected by various complicated physiological disease. Since music is of the most preferable medium that has enormous ability to control human emotion and abstract mood, so there is an opportunity to apply music in non-invasive clinical purpose. Therefore, identification of an appropriate class of music for the treatment of some specific type of diseases has become an important topic of research. At this point, a natural question may arise – Are the music signals used to treat one particular type of diseases be synchronized? In this context, the primary task is to make a study on synchronisation of different music signals.

In communication theory, music is a complex composition of some organized sounds. Physically it means that when some vibration comes from a source, our eardrums start to vibrating and impulses or signals are transmitted to the brain where the impulses are selected, organized, and interpreted. The features extraction based on organized impulses is called psycho-acoustic study. In this connection, some scientific parameters like dynamics, timber, rhythm and tonality are used to quantify the reorganization of the aforesaid characteristics [1]. Degree of loudness or softness in music is called dynamics. Many different kinds of sounds can be produced from one instrument. Timber can classify these sounds. In a musical composition, there are some short and long duration of sound and silence in an organized way. Rhythm is defined as the flow of music through time. Rhythm is generally calculated by measuring tempo of the music signal. In this regard, two music signals are synchronized means that their psycho-acoustic measures are relatively equal. However, if the style of playing is different, the psycho-acoustic measures are found to different and hence proper synchronization [2-15] of the music signals is not established. One of main reason is that all aforesaid psycho-acoustic measures are linear in the sense that the measures are based on linear transforms in frequency domain or time-frequency domain [16]. Since music signals are nonlinear, the aforesaid linear measures thus reveal only linear properties, known as local analysis or short term analysis of the music signals. This short term analysis may not be reliable to establish synchronization [2-15] between two nonlinear signals. In this concern, study of synchronization [2-15] of long term dynamics in term of reconstructed phase spaces [17-23] of two music signals may give some more valuable information.

Long term dynamics of a system are reflected from the phase spaces [17-23] of the signal which is possible to reconstruct [24, 25] with suitable time-delay and proper embedding dimension [26-28]. Whenever the signal is identified as nonlinear by surrogate data test [29-34], the suitable time-delay is generally calculated by Average Mutual Information (AMI) [35-36]. The proper embedding dimension is calculated by False nearest neighbourhood (FNN) method [26-28]. The joining path of the successive embedding points in the phase spaces [17-23] is called trajectory. The different natures of the trajectories in the phase spaces [17-23] reveal different kind of dynamics. Whenever trajectory moves in a bounded region with an erratic way, the system possesses chaos. In this concern, Largest Lyapunov Exponent (LLE) [37-39] is one of the promising tools which assure the existence of chaos in the reconstructed phase spaces [17-23].

In most of the experimental cases, the chaotic phase spaces [17-23] appear very strange in the sense that they contain too much unstable periodic orbits. It is very difficult to describe the movements of the trajectories because of their eccentric behaviour. Recurrence plot (RP) [40-52], a 2D graphical tool is one of the techniques, which describe such movements of the trajectory by only decoding the phase spaces [17-23] points into binary image. The main idea is that when two points in a phase spaces [17-23] fall in a neighbourhood, they are represented by the number '1' and otherwise by '0'. In this way, all points in phase spaces [17-23] are converted into binary matrix or a matrix of binary numbers – 1 & 0. The image of that matrix reveals a 2D diagram which consists of only two colours – black & white. Thus a black point indicates the positions where two trajectories come closer or in other words they are recurrent. In RP [29-34], diagonal line explains how two trajectories are parallel and horizontal lines or vertical lines describe how they are trapped in a position [40-43]. In fact, the periodic trajectories are indicated by these diagonal lines in RP [29-34]. Moreover, based on these parallel lines and horizontal lines some quantification measures are proposed by Norbert Marwan, M. Carmen Romano, Marco Thiel, Jurgen Kurths in 2007 [43,44,52], which quantifies the phase spaces [17-23] more precisely. For the present purpose, the only measure recurrence rate (RR) [44] is used among the all. RR [44] describes how many points are recurrent in a phase spaces [17-23], that means it calculates recurrence density of a phase spaces [17-23].

Many different types of synchronization can be observed in coupled complex systems –Complete Synchronization (CS), Generalized Synchronization (GS), Phase Synchronization (PS) [44, 53-62] etc. When the trajectories evolve due to coupling among themselves, they are known as synchronized completely [44, 53-62]. If any functional relationship exists between the driving and the response systems, they are called synchronized generally. The PS [44, 53-62] of two systems means that their phases adapt to each other, which means that they evolve in the same manner. In this concern, CS is the first study made on chaotic systems. However, only coupled identical chaotic dynamics with different initial conditions evolve onto same trajectory [2, 61, 62]. But this is very rare in experimental cases, because parameters of the system usually mismatch. Thus there is a basic need to study synchronisation between non-identical chaotic systems. Again, increasing the coupling strength of uncoupled non-identical oscillators, a weak degree of synchronization [2-15] may occur in which phases and frequencies of the chaotic oscillators become locked though their amplitudes often remain almost uncorrelated. This fact is generally described in term of PS [44, 53-62]. Moreover, if two systems are synchronized then obviously their recurrences are dependent to each other. Thus, it might be possible to get some indication regarding the degree of their synchronization [2-15] by comparing the recurrences of individual system with joint recurrences of the whole. Further, RP [29-34] reveals recurrences in different time scales which take an important role in explain the dynamics of a system. In fact, for a periodic trajectory, RP [29-34] shows some uninterrupted diagonal lines which are equally spaced. The distances between these diagonal lines are the periods of the trajectory. Hence, there is a relation between the distances of diagonals and the time scales of the system. Moreover, ‡ -recurrence rate [44] is a measured (based on these diagonal lines) which can quantify the degree of PS [44, 53-62]. It ensures us to analyze PS [44, 53-62] by means of ‡ -recurrence [44] of the corresponding reconstructed phase spaces [17-23] of two non identical systems, where phases or time scales are locked.

In this article, we examine both psycho-acoustic and PS [60-62] methods on two instrumental music signals, which are based on same 'raga' played by two different eminent sitar instrumentalist. The first study reveals only those features which are able to distinguish two signals properly with respect to their style and pattern of playing. Moreover in this study, we have seen their scales of notes are completely different. As a result psycho-acoustic features fail to describe synchronization of two aforesaid signals. On the other hand, reconstructed attractors of two music signals possess chaos and different RP [29-34]. But these non identical systems correspond high correlation between their normalized ‡ -recurrence rates [44]. In this way, it is established that their long term dynamics are synchronized in the sense of PS [60-62].

## 2. Methods of Psycho-acoustic study of music signals

All measures related to this section are calculated in MATLAB software under an audio analysing toolbox. The followings are short preview of the methodologies used for such measures.

*Dynamics* measure calculated how soft or loud a music signal. Global energy of any signal can be computed by root-mean-square (RMS). This energy curve shows the temporal distribution of energy. To estimate the low energy rate, the percentage of frames showing less than average energy is calculated. This percentage is actually the degree of loudness or softness.

One common way of describing *timbre* is based on MFCC's [63]. By using the mirspectrum function [5], we can recognize the key in a signal with respect to its corresponding time. The curve containing the key information is called miron set curve [63].Then by calculating slope of the curve, we can get attack slopes for every keys in a signal. In this concern, timber is defined by attack slope and this measure is capable to classify two sounds from same instruments.

*Rhythm* is defined by the measure tempo. After onset detection [63], the periodicity is calculated by autocorrelation for frame decomposition. In order to recognize the periodicities that are more perceptible, the periodogram is filtered using a resonance curve [63], after which the best tempos are estimated through peak picking, and the results are converted into beat per minutes. It is known as tempo of music.

*Tonality* is calculated by key clarity of an audio signal. In this concern, wrapped chromagram [64] is constructed, which shows a distribution of the energy with respect to the twelve possible pitch classes [65]. Krumhansl and Schmuckler [66] proposed a method for estimating the tonality of a musical piece. The most relevent tonality is considered to be the tonality candidate with highest correlation which is known as key strength. Key clarity is defined by key strength associated with the best key(s).

## 3. Phase synchronization of nonlinear signals by means of recurrences

### 3.1. Test of nonlinearity-by surrogate data method

Surrogate data method, initially introduced by Theiler et al. [29-34], which investigate the existence of nonlinear dynamics underlying experimental data. This method proposed a null hypothesis for a specific process class and then compares the system output to this hypothesis. In order to test a null hypothesis at a level of significance $r$, one has to generate $(2/r-1)-1/r$ surrogates for a one side (two-side) test. Then, consider a suitable statistic and compares its value of data to the same of the surrogates. If the value of the statistic of the data deviates from that of the surrogates, then the null hypothesis may be rejected. Otherwise, it may not. In this concern, Diego L. Guar´ın L´opeza et al. present a new surrogate data method [29-34] which is most eminent to test nonlinearity for all type of signals (stationary and non-stationary). This method has three steps – a) Generate 99 surrogate data by Amplitude Adjusted Truncated Fourier Transform (AATFT) [34] from the observed data, b) select the nonlinear version of autocorrelation statistics-AMI (with m=1) as discriminant statistics, c) consider a null hypothesis against which observations are tested. In this connection, we consider the mull hypothesis ($H_0$) as $H_0 : AMI_{\text{experimental signal}}(m=1) = AMI_{SUR(\text{experimental signal})}(m=1)$. AMI [35-36] plays a role as the discriminating statistic, which is basically a number which quantifies some aspect of the time series. If this number is different for the observed data then it would be expected under the null hypothesis, then the null hypothesis can be rejected with level of significance $r = 0.01$.

## 3.2. Average Mutual Information for suitable time-delay

It is an information theoretic approach, which provides formalism for quantifying the concepts of spreading and new information, is used. The most important contributions towards the development and application of MI were made by Fraser and Swinney [35-36]. In this context, Average mutual information (AMI) [35-36] is a particular type of MI between the signal itself.

For a signal $\{x(k), k = 1, 2, ..., N\}$, AMI [35-36] is calculated by

$$AMI(\tau) = \sum_{k=1}^{N-\tau} prob[x(k), x(k+m)] \log \frac{prob[x(k), x(k+m)]}{prob[x(k)] prob[x(k+m)]}, \quad (1.1)$$

$[m = 1, 2, ..., N-1]$ where $prob[\bullet]$ denotes the probability.

For estimation of $\tau$, two criteria are important. First, $\tau$ has to be large enough so that the AMI [35-36] at time $k + \tau$ is significantly different from the AMI [35-36] at time $k$. Second, $\tau$ should not be larger than the typical time for which the system loses memory of its initial state. We will always conscious about the second criteria, because chaotic systems are unpredictable or lose memory of its initial state as time goes forward. In this context, Fraser and Swinney proposed a very useful method which state that the optimum time-delay is obtained where the mutual information attains its first minimum value.

Since, AMI [35-36] is a nonlinear version of autocorrelation, so this AMI [35-36] method is strongly recommended to apply on nonlinear signals for finding suitable time-delay.

## 3.3. Phase space reconstruction

Taken theorem states that it is possible to reconstruct a topological equivalent phase spaces [17-23] from a single time series [24, 25], if the suitable time delay and proper embedding dimension [26-28] can be found out. Let us consider the time series data given by $\{x(k), k = 1, 2, ..., N\}$. Suppose the embedding dimension [26-28] and the delay time for reconstruction of the attractor are $m$ and $\tau$ respectively. Then reconstructed phase spaces [17-23] is given by
$X(k) = (x(k), x(k+\tau), x(k+2\tau), ..., x(k+(m-1)\tau))$, $k = 1, 2, ..., M$, where $\{x(k)\}$ is the phase space's point in $m$-dimension phase spaces [17-23], $M$ is the number of phase points, $M = N - (m-1)\tau$, describes the evaluative trajectory of the system in the phase spaces [17-23]. Reconstruction of the attractor is guaranteed if the dimension of the phase spaces [17-23] is sufficient to unfold the attractor. It is ensured when $m > 2d + 1$ holds, where $d$ is the dimension of the attractor.

For the nonlinear signal, suitable time delay is calculated by the method of AMI [35-36]. The mutual information is a measure of how much information can be derived from one point of a time series, given complete information about the other. On the other hand, embedding dimension [26-28] is calculated by FNN, but for the sake of geometrical visibility we reconstruct the attractor with embedding dimension [26-28] three.

## 3.4. Largest Lyapunov Exponent

Detecting the presence of chaos in a dynamical system is an important problem that is solved by measuring the LLE [37-39]. Lyapunov exponents [37-39] quantify the exponential divergence of initially close state-space trajectories and estimate the amount of chaos in a system. In this context, after reconstructing the attractor the nearest neighbour of each point is located. The nearest neighbours, $X_{\hat{j}}$, is found for such points which minimize the distance to the point $X_j$, known as reference point. This fact is expressed by $d_j(0) = \min_{X_{\hat{j}}} \|X_j - X_{\hat{j}}\|$, where $d_j(0)$ is the initial distance from the *j-th* point to its nearest neighbour and $\|\cdot\|$ denotes the Euclidean norm. Here, the additional constraint is that nearest neighbour have a temporal separation greater than the mean period of the time series, i.e; $|j - \hat{j}| >$ mean period. Mean period is estimated by the reciprocal of the mean frequency of the power spectrum or the median frequency of the magnitude spectrum. To consider each pair of neighbours as nearby initial condition for different trajectories, this method is used. Then LLE ($\lambda_1$)[37-39] is estimated as the mean rate of separation of the nearest neighbours. Since Lyapunov exponent [37-39] quantify the exponential divergence of initial close state-space trajectories, so $d_j(i) \approx C_j e^{\lambda_1(i.\Delta t)}$, where $d_j(i)$ is the distance between the *j*-th pair of nearest neighbours after *i* discrete-time steps, i.e., $i.\Delta t$ seconds, $C_j$ is the initial separation.

The LLE [37-39] is easily calculates by the LLE [37-39] is easily calculated using a least-square fit to the "average" line defined by 
$$y(i) = \frac{1}{\Delta t}\langle \ln d_j(i) \rangle, \qquad (1.2)$$

where $\langle \cdot \rangle$ denotes the average over all values of $j$.

### 3.5. Recurrence plot

Let $x_i \in \mathbb{R}^n$ be any point on the trajectory in an *n*-D phase spaces [17-23]. Then, RP [29-34] is a 2D diagram constructed by a matrix $(R_{ij})_{n \times n}$, where, $R_{ij} = \Theta(\|x_i - x_j\| - \nu)$, $\|\cdot\|$ is the Euclidean norm in $\mathbb{R}^n$ and $\Theta$ is Heaviside function. In other words,

$$R_{ij} = \begin{cases} 1, & \text{if } x_i = x_j, \\ 0, & \text{if } x_i \neq x_j. \end{cases}$$

$\nu$ is chosen as $0.1\sigma$, where $\sigma$ are standard deviation of the time series [67,68].

In 2D diagram, consider '1' as a 'black' dot and '0' as 'white' dot. Therefore, recurrence of any two points in *n*-dimensional phase spaces [17-23] is visualized as a 'black' spot in the 2D diagram. The main diagonal is called line of identity (LOI), which corresponds the recurrence between the points itself.

Recurrence is a fundamental property of a dissipative dynamical system. This tool enables us to investigate the *n*-D phase spaces [17-23] trajectory through a 2-D representation of its recurrences. The initial purpose of RP [29-34] is the visual inspection of higher dimensional trajectories. In RP [29-34], periodic orbits are recognized by a rectangular region made by the diagonal lines. So, when more than two rectangular regions occur in RP [29-34], we can conclude that chaos occurs due to unstable periodic orbits. The diagonal lines in RP [29-34], represent the parallel movements of the trajectories. On the other hand the vertical/horizontal lines represents the trapping time of the trajectories. The rectangular region, formed by diagonal lines, signifies the periodic orbit. In this concern, recurrence rate (RR) [44] is defined by

$$RR(\nu) = \frac{1}{N^2} \sum_{i=1}^{N} R_{i,j}(\nu). \qquad (1.3)$$

This measure is computed for each diagonal line which is parallel to LOI. Now $\ell$-recurrence rate [44] is calculated over all diagonal lines which are parallel to the diagonal line whose distance is $\ell$ from LOI, which is known as $\ell$-recurrence rate [44]. The $\ell$-recurrence rate [44] is thus defined by

$$RR_\ell = \frac{1}{N - \ell} \sum_{i=1}^{N-\ell} R_{i,i+\ell}. \qquad (1.4)$$

## 3.6. Detection of Phase Synchronization by means of $\tau$-recurrences

Two non identical systems meets in PS [44, 53-62] imply their distances between diagonal lines in RP [29-34] are coinciding, because their time scales adapt to each other. Now, if probability of first system recurs after $\tau$ time steps is large, then same of the second system recurs after the same time interval will be also large. Consider $p(\nu,\tau)$ be the probability that a system recurs to $\nu$-neighbourhood of a former point $x_i$ of the trajectory after $\tau$ time steps. Then by comparing probability $p(\nu,\tau)$ of the two systems, it is possible to quantify PS [44, 53-62] clearly. The probability $p(\nu,\tau)$ can be estimated directly from the RP [40-52] by

$$p(\nu,\tau) = RR_\tau(\nu) = \frac{1}{N-\tau}\sum_{i=1}^{N-\tau} R_{i,i+\tau}. \quad (1.5)$$

Consider $RR^x_\tau = \frac{1}{N-\tau}\sum_{i=1}^{N-\tau} R^x_{i,i+\tau}$ and $RR^y_\tau = \frac{1}{N-\tau}\sum_{i=1}^{N-\tau} R^y_{i,i+\tau}$ as two $\tau$-recurrence rates [44] for two phase spaces [17-23].

Then, correlation coefficient between $\overline{RR}^x_\tau$ and $\overline{RR}^y_\tau$ is defined as

$$CPR = \left\langle \overline{RR}^x_\tau \, \overline{RR}^y_\tau \right\rangle, \quad (1.6)$$

where $\overline{RR}^x_\tau$ and $\overline{RR}^y_\tau$ are the probabilities normalised to zero mean and standard deviation one. If $CPR \approx 1$, then both systems are in PS [44, 53-62] that means the probability of recurrence will be maximal at the same time. Otherwise, the optimized probability of recurrence will not occur simultaneously.

## 4. Collection of music signals and experimental results

Two instrumental music signals (based on raga "Malkunse") are collected from internet in mp3 format. Short descriptions of those signals are as follows:

<u>Music signal-1</u>: Classical Background – raga Malkunse, Instrument – Sitar, Performer Name – Pandit Nikhil Banerjee, Sample size – 100000.

<u>Music signal-2</u>: Classical Background – raga Malkunse, Instrument – Sitar, Performer Name – Pandit Ravishankar, Sample size – 100000.

In order to make the analysis meaningful, all signals are segmented from same portion of the raga in same time interval.

### 4.1. Psycho-acoustic study

After calculating RMS energy [63-66], we calculate percentage of window which contains lower energy of the signals. Percentage of lower energy contained window greater means music signal contains very low amplitudes. So, the signal is very soft in listening.

RMS energies of Music signal-1 & Music signal-2 are given in Fig.1a and Fig.1b.

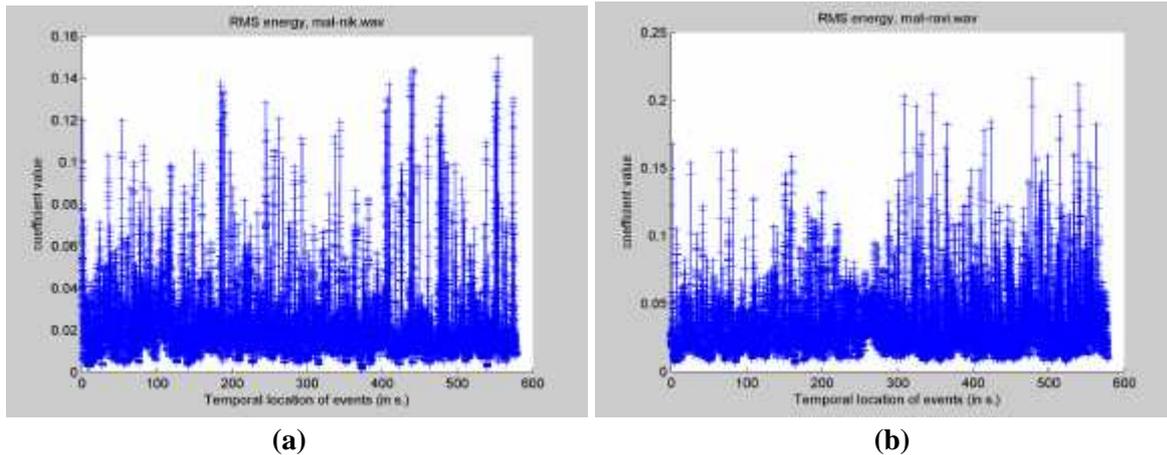

**(a)** **(b)**

Fig.1. RMS energy of (a) music signal-1, (b) music signal-2.

Percentages of lower energy window are shown in table-1.

| Signal | % of lower energy window |
|---|---|
| **Music signal-1** | 0.64831 |
| **music signal-2** | 0.61306 |

Table.1. % of lower energy window of music signal-2 is greater than same of music signal-1

From Table-1, it is implied that loudness of Music signal-1 is less than loudness of Music signal-2 or in other words softness of Music signal-1 is greater than softness of Music signal-2.

If a note has high attack value [63-66], it means the attack is gradual attack otherwise it is steep attack. The attack slopes [63-66] profile of two music signals are given in Fig.2a and Fig.2b.

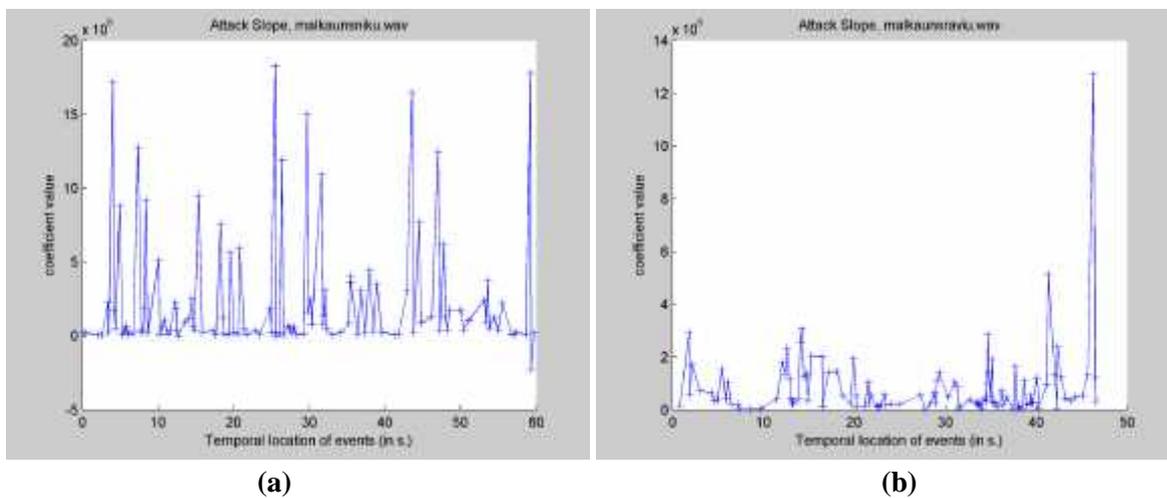

**(a)** **(b)**

Fig.2. Attack slopes profile (in different temporal location of keys) for (a) music signal-1, (b) music signal-2.

In the above figures, we have shown the attack slope [63-66] of each successive note, which indicates whether each note has a steep attack or gradual attack [63-66]. It is observed that, large amount of high attack slope appeared in most of the cases for Music signal-1. On the other hand, in Music signal-2, we observed that most of the attack slope has low values. Thus, sounds of notes Music signal-2 are more aggressive than same of Music signal-1.

Tempo [63-66] is the speed of the beat. A fast tempo in music gives a feeling of excitement and energy, whereas slow tempo [63-66] gives a feeling of calm mood. Tempo [63-66] of Music signal-1 and Music signal-2 are calculated by following way:

we first find the lag where for the first time autocorrelation meets maximum value, then with the corresponding lag we calculate the time where envelope autocorrelation [63-66] attains its maximum and at the end we find bpm for that time. This is the actual value of tempo [63-66] of the corresponding signal.

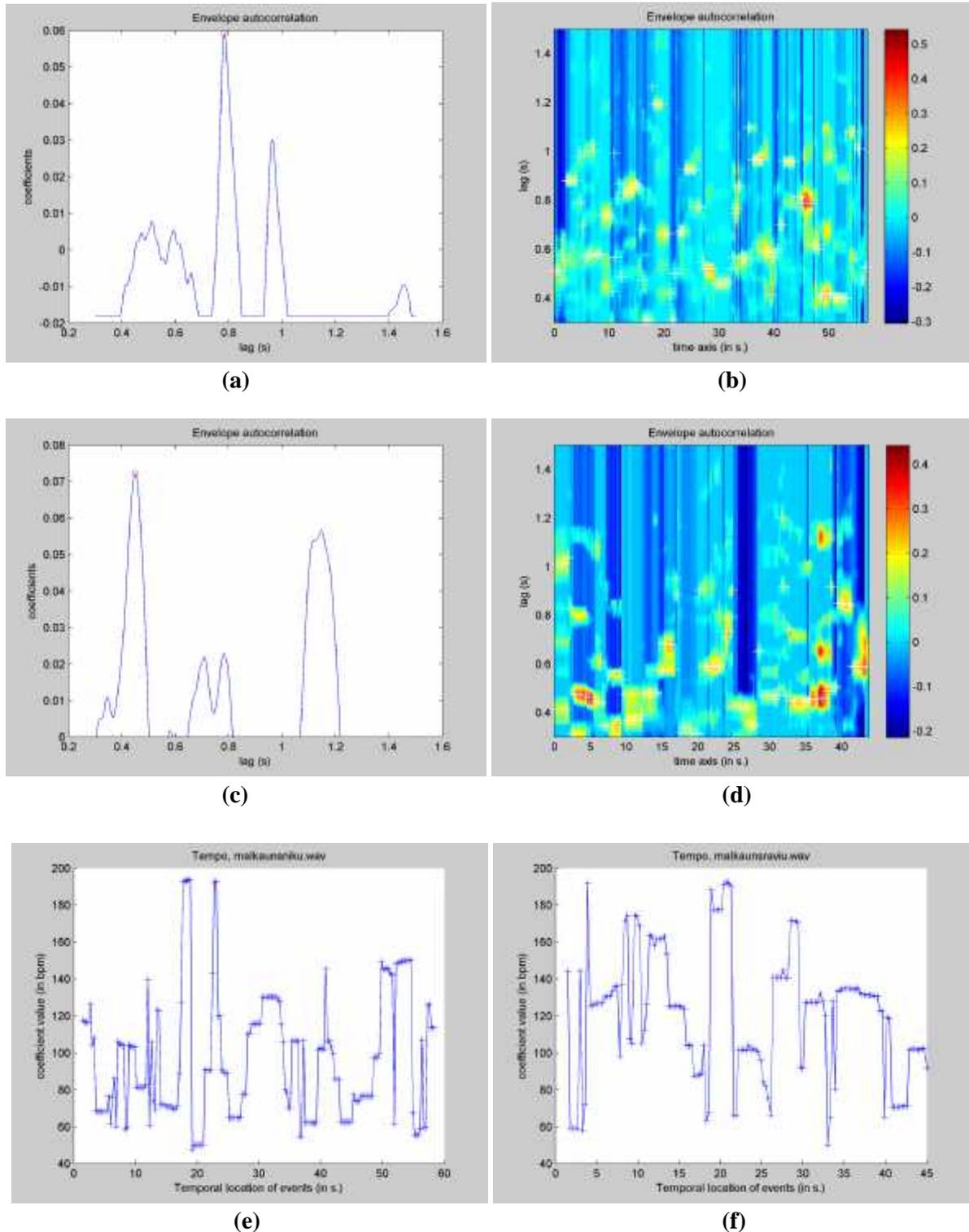

Fig.3.Tempo analysis of Music signal-2.

The results are shown in table.2.

| Signal | Tempo |
|---|---|
| Music signal-1 | 62.71bmp |
| Music signal-2 | 132.9bmp |

Table.2. Tempo of music signal-1 is less than music signal-2

Hence, Music signal-1 is more in calm mood than Music signal-2.

Key clarity [63-66] of the Music signal-1 and Music signal-2 are shown in Fig.4.

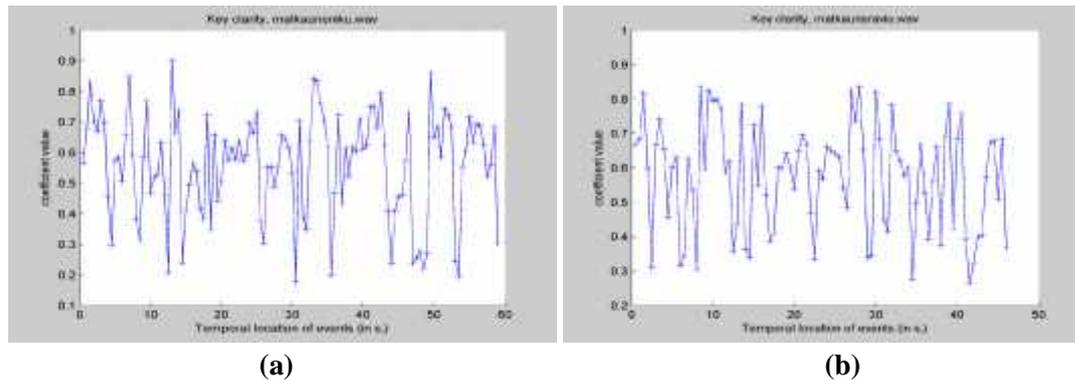

(a)                  (b)

Fig.4. Key clarity of (a) music signal-1, (b) music signal-2.

From the above figures, it is observed that-a) high values lie between 0.8-0.9 for Music signal-1 & Music signal-2; b) low values lie between 0.2-0.3 for Music signal-1 & Music signal-2. So, Clarity of each key for both Music signal-1 & Music signal-2 are approximately same.

### 4.2. Long term phase space analysis of two nonlinear music signals

### 4.2.1. Nonlinearity test of music signal

After generating 99 surrogate data of the music signal-1 & music signa-2, we calculate *AMI* (with *m=1*) for each of them and then draw a graph Grade vs. *AMI* (with *m=1*), which is given in Fig.5 & Fig.7 respectively.

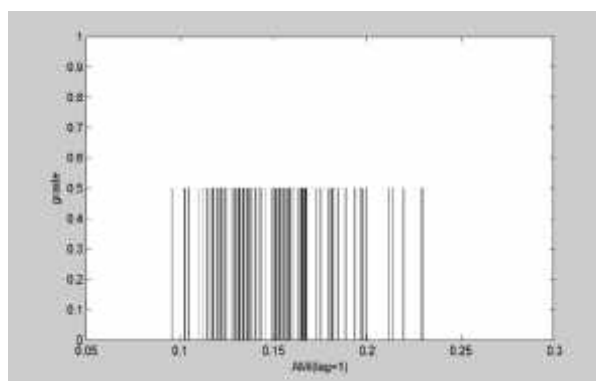

Fig.5. Two grades- '1' and '0' are considered as limit in y-axis. In y-axis, we fix the grade '1' for the music signal-1 and 0.5 for the surrogate data. In x-axis, we take the value of the *AMI* (with lag=*1*).

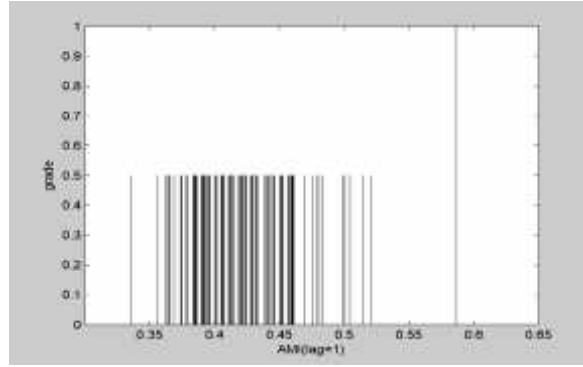

Fig.6.Two grades- '1' and '0' are considered as limit in y-axis. In y-axis, we fix the grade '1' for the music signal-1 and 0.5 for the surrogate data. In x-axis, we take the value of the *AMI* (with lag=*1*).

From Fig.5 & Fig.6, it is seen that value of *AMI* (with *m=1*) of music signal-1 & music signal-2 are different from the values of the *AMI* (with lag=*1*) of each respective 99 surrogate data.

We consider two null hypotheses ($H_0$) as follows:

For music signa-1, $H_{0\text{music signal-1}}$ : $AMI_{\text{music signal-1}}(m=1) = AMI_{\text{Surrogate(music signal-1)}}(m=1)$ and for music signal-2, $H_{0\text{music signal-2}}$ : $AMI_{\text{music signal-2}}(m=1) = AMI_{\text{Surrogate(music signal-2)}}(m=1)$, where $AMI_{\text{music signal-1}}$ denotes *AMI [35-36]* of music signal-1 and $AMI_{\text{Surrogate(music signal-1)}}$ denotes *AMI [35-36]* of surrogate data of music signal-1. The same are defined for music signal-2. Since null hypotheses are rejected, so signals are nonlinear with 0.01% significant level.

### 4.2.2. 3D attractor reconstruction

To reconstruct 3D attractor from music signal-1 and Music signal-2, the suitable time-delays are calculated by the method of AMI [35-36]. AMI [35-36] is a method based on information theory and calculated how much information shared a time series to another. Calculations of AMIs for both the signals are given in Fig.7 & Fig.8.

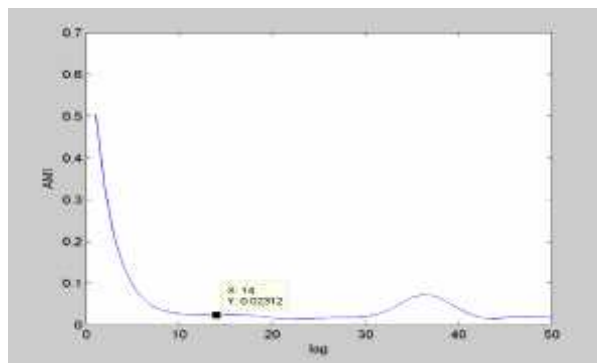

Fig.7. AMI vs. time-delay of the music signal-1. The first minimum is found for lag=14.

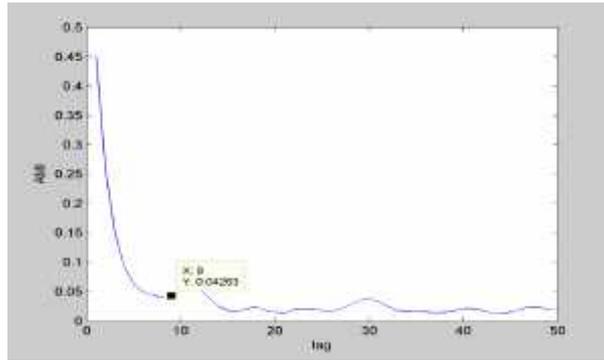
Fig.8. AMI vs. time-delay of the music signal-2. The first minimum is found for lag=9.

The time-delay for music signal-1 is found to be 14 and the same for music signal-2 is found to be 9. With help of method of phase space [17-23] reconstruction (described in section.), the 3D phase spaces of music signal is thus reconstructed. The reconstructed 3D attractors of both the music signals are given in following figure:

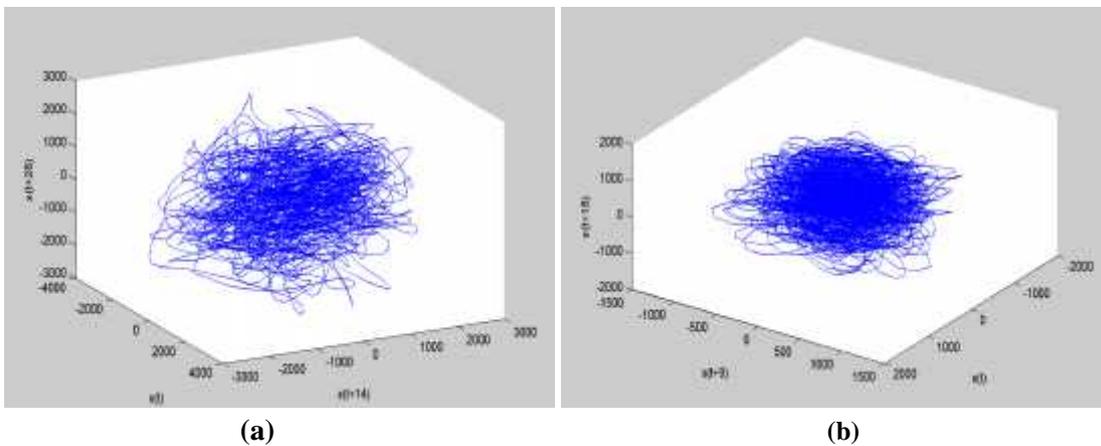

                **(a)**                              **(b)**

Fig.9. 3D reconstructed attractors of (a) music signal-1 with time-delay 14, (b) Music signal-2 with time-delay 9.

### 4.2.3. Quantifying Chaos by Largest Lyapunov Exponent (LLE)

A fitted straight line on the average of log(divergence) of music signal-1 and music signal-2 are given in Fig.10a and 10b.

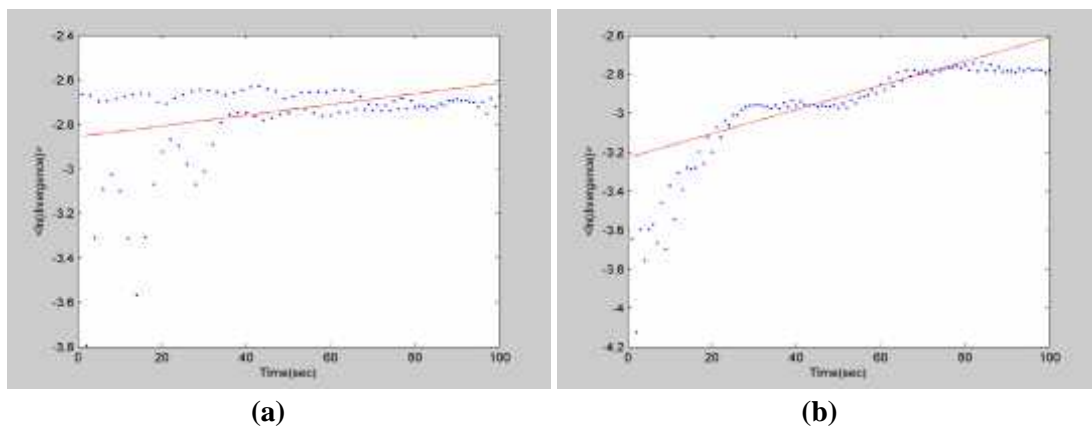

                **(a)**                                **(b)**

Fig.10. Linear straight lines fitted on the linear region on <ln(divergence)> for (a) music signal-1 and (b) music signal-2.

The values of coefficients a & b of the fitted straight lines $y = ax + b$ are given by

$a = 0.0024, b = -2.8550$ (for music signal-1) and $a = 0.0061, b = -3.2276$ (for music signal-2).

Thus values of LLE [37-39] for music signal-1 and music signal-2 are 0.48 and 1.23 respectively. (Sampling frequency is considered 200 Hz).

### 4.2.4. Recurrence Plot

Recurrence plots (RP) [29-34] of music signal-1 with suitable time-delay 14 and embedding dimension 3 is shown in Fig.11a. Also Recurrence plots of music signal-2 with suitable time-delay 9 and embedding dimension 3 is shown in Fig.11b.

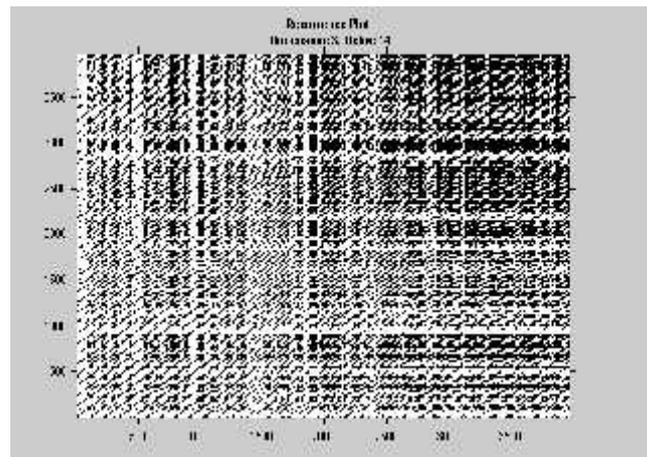

(a)

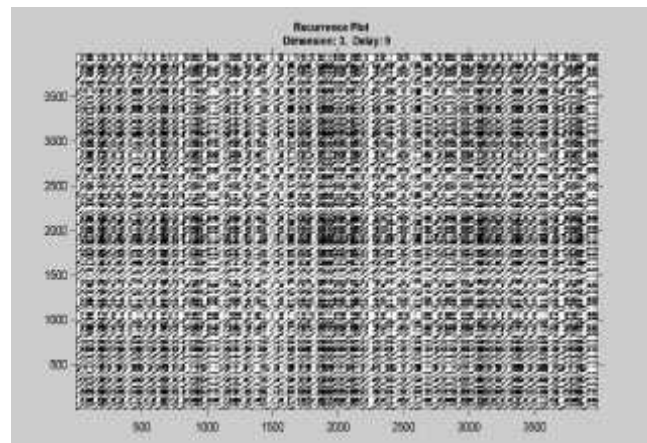

(b)

Fig.11. RP of 3D reconstructed phase spaces [17-23] from (a) Music signal-1 with time-delay 14, (b) Music signal-2 with time-delay 9.

From Fig.11a & Fig.11b, it is observed that RP [29-34] of the 3D reconstructed phase spaces of music signal-1 and music signal-2 are completely different. But there exist many diagonal lines parallel to LOI, which are sometimes occurring in same time. This indicates that there may exists one to one correspondence between phases of two music signals with respect to their ‡-recurrence [44].

**4.2.5. Analysis of ‡-recurrence rate**

It is known in advance that ‡-recurrence rate [44] is higher order correlation between the points of the trajectories with ‡-lag [44]. So, to estimate ‡-recurrence rate of the phase spaces [17-23] with different values of ‡, we get a time series which reflect higher order correlation of the phase spaces [17-23] points. Then, by using the formula (1.5), we can estimate how the phases are similar in different time in the reconstructed phase spaces [17-23].

‡-recurrence rate [44] of Music signal-1 and Music signal-2 are given in following figure:

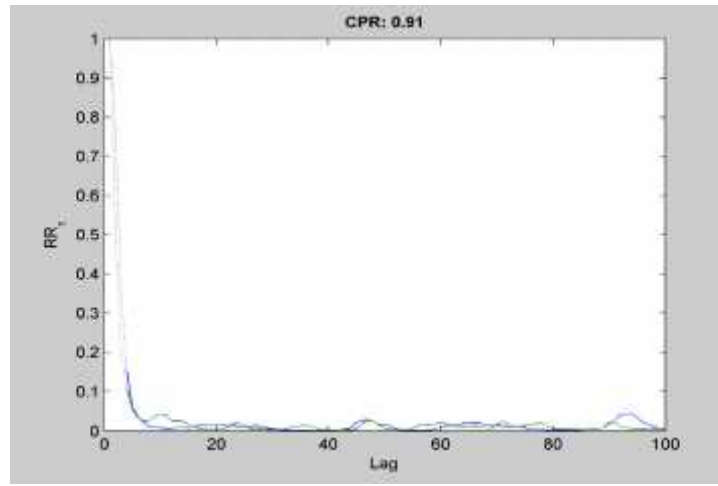

Fig.12. Values ‡-recurrence rate of Music signal-1 and Music signal-2 are calculated for 100 values of ‡.

By applying the formula (1.6), it is seen that value of CPR=0.91. So, degree of PS of Music signal-1 and Music signal-2 is very high. Thus phases of aforesaid music signals are well synchronized.

## 4. Conclusions

This article produces two different studies on two music signals to understand music more vividly. The first study viz. psycho-acoustic study reveals how two signals are different with their styles of playing. The measure dynamics shows music signal-1 is softer than music signal-2. Timber analysis possesses notes of sound in music signal-2 is aggressive than music signal-1. Analysis of tempo reveal music signal-1 is more in calm mood than music signal-2. At the end of this study, tonality shows that the key clarities of both the signals are approximately same. However, from the above features, it is not possible to find any similarity between the two signals, except that they are played under same raga on same instrument as we know before. Naturally, it is a challenge to extract some more information from those signals and established the similarity between them. In fact, communication theory in music signal demands a study which can establish similarity between two signals with respect to some informative knowledge. In this concern, method of nonlinear dynamics and concept of phase synchronization [60-62] gives such aspect which is analyzed in this study. With concise view, it is seen that the phase of their 3D reconstructed attractor are approximately similar with respect to their recurrence rate of chaotic phase spaces [17-23] points in different scales. Also high value of correlation of the normalized ‡-recurrence [44] is 0.91 which corresponds two non identical oscillators is synchronized with respect to their phases. Thus, phase synchronization [60-62] is one of the quantification tools in communication theory to analyze the similarity of two music signals.